\documentclass{aanew}
\usepackage{epsf}

\begin{document}

   \title{Discovery of circular polarization in the Intermediate Polar 1WGA
   J1958.2+3232\thanks{Based on observations made with the Nordic
   Optical Telescope, operated on the island of La Palma jointly
   by Denmark, Finland, Iceland, Norway and Sweden, in the Spanish
   Observatorio del Roque de los Muchachos of the Instituto de
   Astrofisica de Canarias.}}

   \author{M. Uslenghi\inst{1}
          \and
          L. Tommasi\inst{2}
          \and
          A. Treves\inst{2}
          \and
          V. Piirola\inst{3}
          \and
          P. Reig\inst{4,5}
          }

\offprints{uslenghi@ifctr.mi.cnr.it}

\institute{ Istituto di Fisica Cosmica "G.Occhialini",CNR, Via
Bassini 15, I-20133 Milano, Italy.
        \and
Universit\`a degli Studi dell'Insubria, Polo di Como,
Via
Valleggio 11, 22100, Como, Italy.
         \and
Tuorla Observatory, SF21500 Piikkio, Finland
             \and
Foundation for Research and Technology - Hellas, GR-710 10
Heraklion, Crete, Greece
\and
Physics Department, University of
Crete, GR-710 03 Heraklion, Crete, Greece
             }

   \date{Received September 00, 0000; accepted March 00, 0000}
   \abstract
   {We report on UBVRI polarimetry of the recently
   identified Intermediate Polar \object{1WGA J1958.2+3232},
  carried out on 2000 August, 4-6 at the Nordic Optical
  Telescope. Circular polarization was detected in R and I bands
  with an increasing absolute mean value with wavelength. There
  is evidence of possible modulation of the circular polarization
  at twice the previously reported white dwarf spin period,
  suggesting that it is the true period and that the modulation
  in optical and X-ray is dominated by the first harmonic.
  Indication of modulation at the orbital period is also present.
  \keywords{stars: individual: \object{1WGA J1958.2+3232} --
                novae, cataclysmic variables --
                binaries: close --
                stars: magnetic fields --
                polarization
               }
   }

%

  \titlerunning{Circular polarization in \object{1WGA J1958.2+3232}}
   \authorrunning{M. Uslenghi et al.}
   \maketitle

\section{Introduction}

\object{1WGA J1958.2+3232} is one of the objects found by Israel
et al. (\cite{Isr98}) in a systematic search for pulsators in the
catalogue of ROSAT X-ray sources compiled by White, Giommi \&
Angelini (\cite{WGA94}). The source appeared modulated with a
period 721 $\pm$ 14 \textrm{s} and with a pulsed fraction of about
80\%. A subsequent observation performed by ASCA confirmed both
the flux level and the strong periodic modulation at 734 $\pm$ 1
\textrm{s} (G.L. Israel 2000, private communication), which was
interpreted as a spin period.

On the basis of optical photometry and low resolution
spectroscopy, Israel et al. (\cite{Isr99}) proposed a $m_V=15.7$
star as the optical counterpart of  \object{1WGA J1958.2+3232}.

Further spectroscopic measurements carried out by Negueruela et
al. (\cite{Neg99}), as well as time resolved optical photometry
performed by Uslenghi et al. (\cite{Usl00}) (which provided
detection of optical modulation at $733.24\pm0.02$~\textrm{s}),
confirmed the candidate of Israel et al. (\cite{Isr99}) and
allowed its classification as an Intermediate Polar (IP).

A 4.36$^\textrm{h}$ modulation was also detected by Zharikov et
al. (\cite{Zha00}), in time-resolved photometric and spectroscopic
observations, and interpreted as due to the orbital period.

We report here on the results of the first polarimetric
measurements on this object.

\section{Observations}
The observations were carried out at the 2.56 \textrm{m} Nordic
Optical Telescope (NOT), La Palma, equipped with the TURPOL
double image chopping photopolarimeter (Piirola \cite{Pii73},
\cite{Pii88}). The instrument allows simultaneous measurements in
the UBVRI bands and was operated in the simultaneous linear and
circular polarization mode.

In this mode, polarization of the light is explored by a
quarter-wave retarder plate, rotated through eight positions by
22.5$^\circ$ steps. A calcite slab then splits the light into
ordinary and extraordinary rays, that pass through identical
diaphragms. Both components of sky background light enter in the
two diaphragms, resulting in a cancellation of the polarization of
sky light. Finally, the beam is split by four dichroic+bandpass
filter combinations and sent to five photomultipliers. Each of
these channels reproduces the spectral response of one of the
\mbox{UBVRI} Johnson-Cousins bands. In this way, truly
simultaneous multiband observations can be performed. For all the
observations, an integration time of 10 seconds was used for each
position of the retarder plate, giving a polarimetric measurement
every 3.5 minutes (whereas the time resolution for photometry is
about 23 \textrm{s}). A 10 seconds sky background integration was
normally performed every 15 minutes. High (HD 15445 and HD
204827) and null (BD +32$^{\circ}$3739 and BD +28$^{\circ}$4211)
polarization standard stars from Schmidt et al. (\cite{Sch92}),
observed many times per night (at the beginning, near midnight
and at the end), were used to determine instrumental polarization
and the orientation of the zero point of the retarder plate with
respect to North.

We monitored \object{1WGA J1958.2+3232} for a total of about 13
hours, during three nights, on 2000 August 4, 5 and 6 (see the
Journal of Observations in Table~\ref{TabOss}). However, due to
poor weather conditions, the signal-to-noise ratio during the
first night is quite low and the data have been discarded.

\begin{table}
      \caption[]{Journal of Observations.}
\begin{tabular}{c c c c}
  \hline
  \hline
  HJD & UT start & UT stop & n.measures \\
  \hline
  2451761$^{\mathrm{a}}$ & 01.03.28 & 03.58.47.& 52 \\
  2451762 & 22.26.27 & 04.13.12 & 76 \\
  2451763 & 22.44.40 & 04.34.04 & 84 \\ \hline
\end{tabular}
\begin{list}{}{}
\small {\item[$^{\mathrm{a}}$] August 4, 2000. }
\end{list}
      \label{TabOss}
   \end{table}

\section{Data analysis and results}
\label{hairymath} Data reduction was performed using dedicated
routines. They allow calculation of $Q$, $U$ and $V$ Stokes
parameters by fitting the counts recorded in the eight positions
of the retarder plate with a suitable cosine function, using a
least-squares algorithm. Subtraction of a mean sky value obtained
by interpolating between the nearest sky background acquisitions
taken close to the object is automatically performed. Finally,
the measured Stokes parameters are corrected by taking into
account the instrumental constants, determined through
observation of standard polarized stars. The error estimate takes
into account both photon statistics and uncertainty of the
least-squares fit.

Table~\ref{PolCir} and \ref{Pollin} report the two night means of
circular and linear polarization, respectively. The former is
characterized by amplitude and sign of the polarization
percentage, the second by polarization percentage and position
angle. Overall means of all the circular polarization collected
data on both nights are reported in the last column of
Table~\ref{PolCir}.

\begin{table}
      \caption[]{Circular Polarization.}
\begin{tabular}{c c c c }
    \hline
    \hline
  Band & 2000 Aug 5 (\%)& 2000 Aug 6 (\%)& Average (\%)\\
  \hline

  $U$ & $-$0.030$\pm$0.108 & $+$0.098$\pm$0.080 & $+$0.038$\pm$0.064 \\
  $B$ & $+$0.078$\pm$0.144 & $+$0.209$\pm$0.104 & $+$0.164$\pm$0.084\\
  $V$ & $+$0.110$\pm$0.170 & $+$0.227$\pm$0.140 & $+$0.179$\pm$0.108 \\
  $R$ & $-$0.535$\pm$0.128 & $-$0.553$\pm$0.106 & $-$0.546$\pm$0.082\\
  $I$ & $-$0.921$\pm$0.197 & $-$0.852$\pm$0.180 & $-$0.910$\pm$0.136\\
  \hline
  \label{PolCir}
\end{tabular}
 \end{table}

\begin{table}
      \caption[]{Linear Polarization.}
\begin{tabular}{c c c c c c c }
\hline \hline
   & \multicolumn{2}{c}{2000 Aug 5} & \multicolumn{2}{c}{2000 Aug 6}  \\
  Band & P (\%) & PA ($^\circ$)& P (\%) & PA ($^\circ$)  \\
   \hline
  $U$& 1.37$\pm$0.22& 141.2$\pm$4.5& 0.88$\pm$0.16& 143.2$\pm$5.2\\
  $B$& 1.53$\pm$0.29& 164.2$\pm$5.3& 0.84$\pm$0.21& 157.5$\pm$7.0\\
  $V$& 1.41$\pm$0.34& 148.1$\pm$6.8& 0.71$\pm$0.28& 114.7$\pm$10.7\\
  $R$& 1.06$\pm$0.26& 151.6$\pm$6.8& 0.97$\pm$0.21& 141.9$\pm$6.2\\
  $I$& 1.31$\pm$0.39&   6.9$\pm$8.4& 0.96$\pm$0.36& 104.6$\pm$10.3\\ \hline
\end{tabular}
\label{Pollin}
 \end{table}

\subsection{Wavelength dependence of circular polarization}

A strong wavelength dependence is apparent, with circular
polarization increasing, in absolute value, toward the infrared.
This is a common result for polarized IPs (\object{BG CMi},
Penning et al. \cite{Pen86}; \object{PQ Gem}, Piirola et al.
\cite{Pii93}; \object{RX J1712.6-2414}, Buckley et al.
\cite{Buc97}) and it is generally attributed to cyclotron
emission, possibly in combination with free-free emission
(Chanmugam \& Frank \cite{Cha87}, West et a. \cite{Wes87},
Piirola et al. \cite{Pii93}, V\"{a}th \cite{Vat97}, Buckley
\cite{Buc00}). Cyclotron radiation produces very large circular
polarization and the emission is strong at low harmonics, but
drops suddenly at higher harmonics. The drop is used to estimate
the strength of the magnetic field: typical values estimated for
the polarized Intermediate Polars are in the range 5-10
\textrm{MG} (see, e.g., Piirola et al. \cite{Pii93}), below the
10-100 \textrm{MG} reported for the Polars.

\begin{figure}[htbp]
  \begin{center}
    \leavevmode
     \epsfxsize=7.5cm \epsfbox{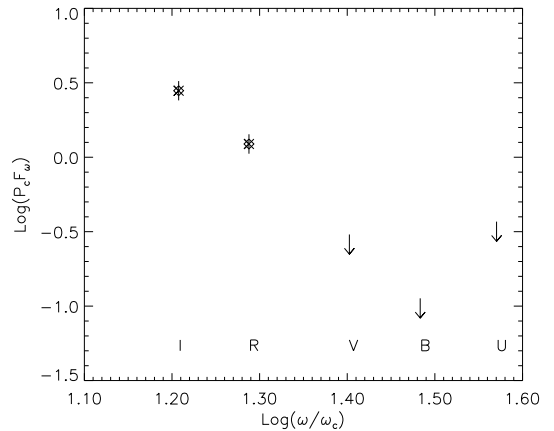}
    \caption{Wavelength dependence of the circularly polarized
    fluxes; $\omega_{\textrm{c}}$ is referred to a 8 \textrm{MG} field. Each point is calculated from the overall
    observations.}
    \label{fig1}
    \end{center}
\end{figure}

In Fig.~\ref{fig1} the observed polarized fluxes
($F_{\textrm{c}}=P_{\textrm{c}}F_{\textrm{tot}}$) are reported for
the $R$ and $I$ bands,together with the 3$\sigma$ upper limits in
\mbox{UBV}, versus $\omega/\omega_{\textrm{c}}$, where
$\omega_{c}$ is the cyclotron frequency for a typical magnetic
field of 8 \textrm{MG}. Normalization is arbitrary for both axis
but, since the scale is logarithmic, changes in normalization
result in a shift only.

\subsection{Polarimetric variability}
Polarimetric data were analysed for the presence of periodic
variations related to the dynamics of the binary system.

\subsubsection{Circular polarization}
\begin{figure}[htbp]
  \begin{center}
    \leavevmode
     \epsfxsize=6.5cm \epsfbox{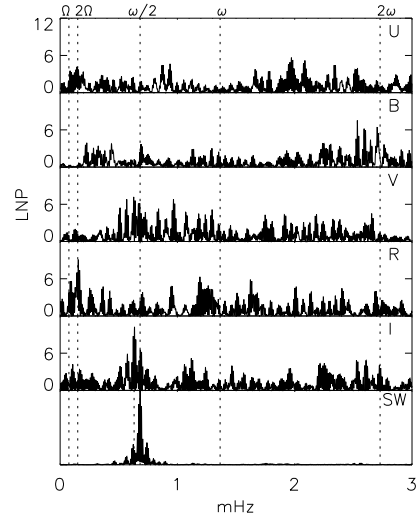} \caption{Lomb-Scargle
    periodogram of the circular polarization data in all bands.
    In the bottom panel, the spectral window is shown (it has
    been shifted in order to have the peak at $\omega$/2).}
    \label{fig6}
    \end{center}
\end{figure}

In Fig.~\ref{fig6} the Lomb-Scargle periodogram (Scargle
\cite{Sca82}) of the degree of the circularly polarized light in
\mbox{UBVRI} bands is reported. $\Omega$ and $\omega$ are the
purported orbital and white dwarf spin frequencies, respectively
($\Omega^{-1}$=4.36\textrm{h}, Zharikov et al. \cite{Zha00},
$\omega^{-1}$=733.2\textrm{s}, Uslenghi et al. \cite{Usl00}). No
evidence of a periodicity at $\omega$ is present. However, some
power around half of this frequency can be seen in all bands
other than $U$. The clearest evidence is in the $I$ filter, where
the second highest peak in the periodogram occurs at
1464$\pm$7~\textrm{s}, consistent with $\omega$/2. Since the
frequency is known \textit{a priori}, in computing the
significance level for this peak the correction for multiple
trials does not apply and the significance level, due to the
exponential distribution of the power at a given frequency, is
simply given by $1-e^{-z}$, where $z$ is the value of the
periodogram (i.e. the power normalized by the variance). In this
case ($z=6.6$) the significance level is 99.8\% (in effect, the
presence of non-white noise results in a significance level lower
than the theoretical one). Moreover, the highest peak is found at
an aliased frequency of $\omega$/2, produced by the sampling
pattern of the polarimeter, as apparent in the spectral window
(bottom panel in Fig.~\ref{fig6}). This raises the possibility
that the true period is 1466~\textrm{s}, while the previously
reported spin periods (e.g. from the X-rays) are the first
harmonic.

\begin{figure}[htbp]
  \begin{center}
    \leavevmode
     \epsfxsize=6.5cm \epsfbox{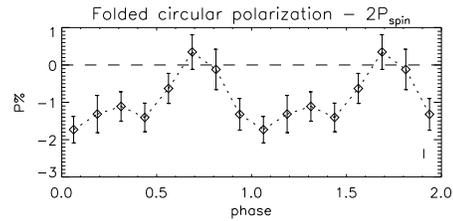} \caption{Circular
    polarization data in $I$ band, folded with the 1466~\textrm{s} period and
    rebinned. The time of the zero phase for all the folded plot
    in this paper is HJD=2451761$^{\textrm{d}}$.2281 (Zharikov et al.
    \cite{Zha00}).}
    \label{fig2}
    \end{center}
\end{figure}

Fig.~\ref{fig2} shows the I-band data folded on the
1466~\textrm{s} period and rebinned. Modulation is apparent, with
peak-to-peak amplitude of about 2\% and mean value of about -1\%.

\begin{figure}[htbp]
  \begin{center}
    \leavevmode
     \epsfxsize=6.5cm \epsfbox{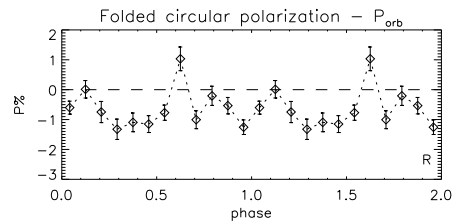}
    \caption{Circular
    polarization data in $R$ band, folded with the orbital period (4.36$^{\textrm{h}}$, Zharikov et al.
    \cite{Zha00}), and rebinned. The time of the zero phase is HJD=2451761$^{\textrm{d}}$.2281.}
    \label{fig3}
    \end{center}
\end{figure}

Modulation at the orbital period and at its first harmonic is
also present in the $R$ band. The data folded over the orbital
period are presented in Fig.~\ref{fig3}.

\subsubsection{Linear polarization}
Due to the reduced efficiency (50~\%) of the linear polarization
measurements in the simultaneous linear and circular polarization
mode, the data have a lower signal to noise ratio than for the
circular polarization. A Lomb-Scargle periodograms of the $Q/I$
and $U/I$ normalized Stokes parameters does not show any
significant peak. Near the frequencies of interest, we carried
out a $\chi^{2}$ epoch folding analysis (Leahy et al.
\cite{Lea83}), which is more sensitive than the Fourier
techniques to non-sinusoidal signals, but no significant features
were apparent.

The relatively flat wavelength dependence of the average linear
polarization (Table~\ref{Pollin}) is within the measurement
errors consistent with that of typical interstellar polarization,
and there is no clear evidence of intrinsic component from the
nightly mean polarization values.

\subsection{Photometry}
\begin{figure}[htbp]
  \begin{center}
    \leavevmode
     \epsfxsize=6.5cm \epsfbox{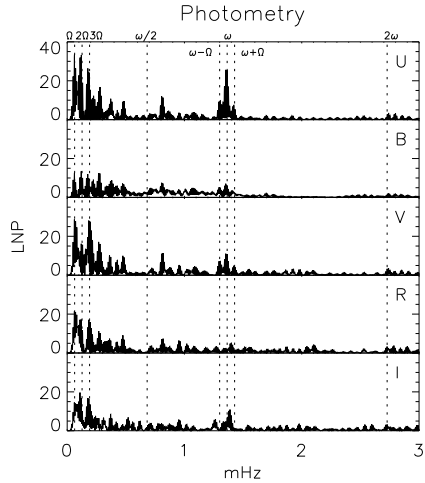}
    \caption{Lomb-Scargle periodogram of the flux in \mbox{UBVRI} bands.}
    \label{fig8}
    \end{center}
\end{figure}

Photometric data were first corrected for atmospheric extinction
and then a 3$^{rd}$ order polynomial was subtracted in order to
remove long time scale variations due to residual atmospheric
effects. Periodograms of the flux intensity in the five bands are
presented in Fig.~\ref{fig8}. In $U$ band a well defined peak
occurs at $\omega^{-1}$=733.2$\pm$0.1~\textrm{s} and synodic
pulses at both $\omega-\Omega$ and $\omega+\Omega$ are clearly
apparent. In the other bands the $\omega$ peak is weaker, but
some power concentration is always present. Modulation at the
orbital period ($\Omega^{-1}$=4.46$\pm$0.1~\textrm{h}, with
1/\textrm{d} alias at 3.76 and 5.48~\textrm{h}) and at its first
and second harmonic is also detected. Instead, we cannot find any
indication of the 1466~\textrm{s} period and light curves folded
at this period show no significant odd-even effect. If the spin
period is indeed 1466~\textrm{s}, both poles must emit in a very
symmetrical manner.

\begin{figure}[htbp]
  \begin{center}
    \leavevmode
     \epsfxsize=6.0cm \epsfbox{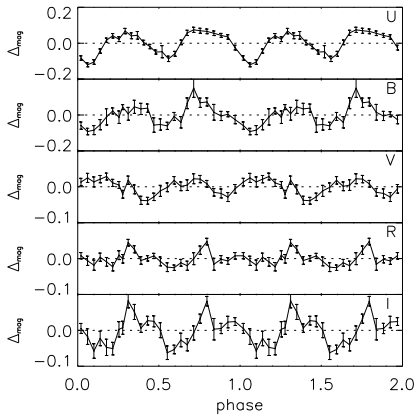} \caption{Folded light
    curves at the 1466~s period with 25 bin, in magnitudes
    relative to the mean level. The standard errors
of the bin values were calculated from the scatter of the
individual integrations (in this case 40).} \label{fig9}
    \end{center}
\end{figure}

\section{Discussion}
Polarization curves and measurements of cyclotron lines are used
to measure the magnetic field and the topology in Polars.
Instead, in Intermediate Polars, which have lower magnetic fields,
polarization is rarely detected. To our knowledge, only four other
IPs show significant polarization: \object{BG CMi} (Penning et
al. \cite{Pen86}), \object{PQ Gem} (Piirola et al. \cite{Pii93}),
\object{RX J1712.6-2414} (Buckley et al. \cite{Buc97}) and
\object{V405 Aur} (Shakhovskoj \& Kolesnikov \cite{Sha97},
Piirola et al. \cite{Pii01}). The circular polarization of
\object{BG Cmi} appears to be constant, whereas the latter three
show spin modulation: in all cases the polarization is in the
order of few \%.

Our data are not sufficient to establish conclusively
phase-dependent variability of the polarization at any of the
characteristic frequencies, but, surprisingly, the clearest
indication is for a modulation at a period twice the accepted
spin period. This seems provide a strong evidence that the spin
period is 1466~\textrm{s} and that the main frequency in optical
and X-ray light curves is its first harmonic. In Uslenghi et al.
\cite{Usl00} evidence for optical flux modulation at the
1466~\textrm{s} period was reported, but for one night only.
However, among the IP, there are other systems ( \object{YY Dra}
and \object{V405 Aur}, Allan et al. \cite{All96}) which have
optical and X-ray light curves dominated by the first harmonic.

\begin{acknowledgements}
We thank the NOT staff for technical support during the
observations, in particular Anlaug Amanda Kaas and Carlos Perez.
We are grateful to Lucio Chiappetti for helpful conversations and
to Santo Catalano for a critical reading of the manuscript and
for his valuable comments. Financial support from EC grant
ERBFM-RXCT 98-0195 and Italian Murst COFIN 98021541 are
acknowledged.

\end{acknowledgements}


\begin{thebibliography}{}

\bibitem[1996] {All96} Allan A., Horne K., Hellier C., Mukai K., Barwig H., Bennie P.J., Hilditch R.W., 1996, MNRAS, 279, 1345-1348
\bibitem[1997] {Buc97} Buckley D.A.H., Haberl F., Motch C., Pollard K., Schwarzenberg-Czerny A., Sekiguchi K., 1997, MNRAS, 287, 117
\bibitem[2000] {Buc00} Buckley, D. A. H., 2000, NewAR, 44, 63
\bibitem[1987] {Cha87} Chanmugam, G., Frank J., 1987, ApJ, 320, 746-755
\bibitem[1998] {Isr98} Israel G.L., Angelini L., Campana S., Giommi P., Stella L., White N.E., 1998, MNRAS, 298, 502-506
\bibitem[1999] {Isr99} Israel G.L., Covino S., Polcaro V.F., Stella L., 1999, A\&A, 345, L1-L4
\bibitem[1983] {Lea83} Leahy D.A., Darbro W., Elsner R.F.,
Weisskopf M.C., 1983, ApJ, 266, 160-170
\bibitem[2000] {Neg99} Negueruela I., Reig P., Clark J.S., 2000, A\&A, 354, L29-L32
\bibitem[1986] {Pen86} Penning W.R., Schmidt G.D., Liebert J.,
1986, ApJ, 301, 881
\bibitem[1973] {Pii73} Piirola V., 1973, A\&A, 27, 383
\bibitem[1988] {Pii88} Piirola V., 1988, in 'Polarized radiation
of circumstellar origin', Coyne G.V. et al. (eds.), University of
Arizona Press, p.735
\bibitem[1993] {Pii93} Piirola V., Hakala P., Coyne G.V., 1993, ApJ,
410, L107-L110
\bibitem[2001] {Pii01} Piirola V. et al., 2001, in preparation
\bibitem[1982] {Sca82} Scargle J.D., 1982, ApJ, 263, 835-853
\bibitem[1992] {Sch92} Schmidt G.D., Elston R., Lupie O.L., 1992,
AJ, 104, 1563
\bibitem[1997] {Sha97} Shakhovskoj N.M., Kolesnikov S.V., IAUC
6760
\bibitem[2000] {Usl00} Uslenghi M., Bergamini P., Catalano S., Tommasi L.,
Treves A., 2000, A\&A, 359, 639
\bibitem[1997] {Vat97} V\"{a}th H., 1997, A\&A, 317, 476
\bibitem[1987] {Wes87} West S.C., Berriman G., Schmidt G.D., 1987,
ApJ, 322, L35-L39
\bibitem[1994] {WGA94} White N.E., Giommi P., Angelini L., 1994, IAU Circ. 6100
\bibitem[2000] {Zha00} Zharikov S.V., Tovmassian G.H., Echevarr$\acute{i}$a
J., C$\acute{a}$rdenas A.A., 2001, A\&A, 366, 834-839
\end{thebibliography}
\end{document}